\documentclass[prb,amsmath,amssymb,superscriptaddress,aps,twocolumn,floatfix]{revtex4-1}

\usepackage{graphicx}
\usepackage{dcolumn}
\usepackage{bm}
\usepackage[caption=false]{subfig}
\usepackage{verbatim}
\begin{document}

\preprint{APS/123-QED}

\title{Intrinsic Magnon Nernst Effects in Pyrochlore Iridate Thin Films}

\author{Bowen Ma}
\affiliation{
 Department of Physics, The University of Texas at Austin, Austin, Texas 78712, USA\\
}

\author{Gregory A. Fiete}
\affiliation{
 Department of Physics, Northeastern University, Boston, Massachusetts 02115, USA\\
}
\affiliation{Department of Physics, Massachusetts Institute of Technology, Cambridge, Massachusetts 02139, USA}

\date{\today}

\begin{abstract}
We theoretically study the magnon spin thermal transport using a strong coupling approach in pyrochlore iridate trilayer thin films grown along the [111] direction.  As a result of the Dzyaloshinskii-Moriya interaction (DMI), the spin configuration of the ground state is an all-in/all-out ordering on neighboring tetrahedra of the pyrochlore lattice. In such a state, the system has an inversion symmetry and a Nernst-type thermal spin current response is well defined. We calculate the temperature dependence of the magnon Nernst response with respect to the magnon band topology controlled by the spin-orbit coupling parameters and observe topologically protected chiral edge modes over a range of parameters. Our study complements prior work on the magnon thermal Hall effect in thin-film pyrochlore iridates and suggests that the [111] grown thin-film pyrochlore iridates are a promising candidate for thermal spin transport and spin caloritronic devices.
\end{abstract}

\maketitle


\section{\label{sec1}Introduction}
In recent years, the entwining of heat and spin transport in the field of spin caloritronics has aroused great interest. Many spin analogies of thermoelectric effects, such as spin Seebeck effects,\cite{uchida2008observation,jaworski2010observation} thermal spin-transfer torques\cite{hatami2007thermal} and spin Nernst effects,\cite{meyer2017observation} have been discovered, leading to the new field of spin caloritronics.\cite{uchida2021transport,bauer2012spin} A collective excitation carrying both energy and spin angular momentum in a magnetically ordered material, magnons often exhibit low dissipation.\cite{nakata2015wiedemann} The utilization of magnons as spin carriers in magnetic insulators has attracted particular attention.\cite{chumak2015magnon} Moreover, the spin-orbit coupling (SOC) in magnetic systems leads to an asymmetric Dzyaloshinskii-Moriya interaction (DMI) which promotes non-collinear magnetic textures.\cite{dzyaloshinsky1958thermodynamic,moriya1960anisotropic} The DMI also enriches the properties of magnon bands and these systems may exhibit Hall-like effects, such as the magnon thermal Hall effect\cite{katsura2010theory,onose2010observation} and the magnon Nernst effect  (MNE).\cite{cheng2016spin,zyuzin2016magnon,shiomi2017experimental}

Among insulating magnetic materials with strong DMI, the pyrochlore family of iridate compounds has garnered significant attention.\cite{kargarian2013topological,witczak2014correlated} In the limit of strong electron-electron interaction, pyrochlore iridates possess a non-collinear antiferromagnetic insulating state in the presence of the DMI.\cite{shinaoka2012noncollinear} These magnetic ground states can provide arbitrarily polarized spin currents and may be switched to other ordered ground states with different magnetic point group symmetries, as well as localized spin textures with nontrivial topology, e.g. skyrmions.\cite{bogdanov2001chiral} Moreover, with spin-orbit coupling and strong electronic correlations, these materials may exhibit novel phases such as axion insulators,\cite{chen2012magnetic,wan2010electronic,go2012correlation} topological Mott insulators\cite{maciejko2014topological,pesin2010mott}, and Dirac or Weyl semi-metals.\cite{witczak2012topological,PhysRevB.84.075129} The metal-to-insulator transition induced by increasing the electron-electron interaction brings additional interesting physics to the fore.\cite{matsuhira2011metal,yanagishima2001metal} These features make the pyrochlore iridates a promising platform for studying the interplay between spin currents, heat flow, and band topology, for both electrons and magnons.

In this paper we study the intrinsic magnon Nernst effect in a trilayer all-in/all-out (AIAO) pyrochlore iridate, with formula A$_2$Ir$_2$O$_7$, in the strong coupling limit with different values of the DMI. Here A is typically a rare earth element, which we will assume has no moment in this work (true for certain choices of A), Ir is iridium and O is oxygen. In Sec.~\ref{sec2}, we introduce the spin model and calculate the topological magnon band evolution with the DMI using a generalized Bogoliubov transformation. We also investigate the spin current chirality of the topologically protected edge modes in a strip geometry. In Sec.~\ref{sec3}, we calculate the spin Berry curvature of a thin film trilayer to study the effects of the DMI on the magnon Nernst response coefficients.  We find sign changes of the coefficients by modifying the DMI or raising the  temperature of the system. In Sec.~\ref{sec4}, provide a discussion on experimental realizations and summarize the main conclusions of our work.

\section{\label{sec2}Model}
\begin{figure}
\includegraphics[width=0.4\textwidth]{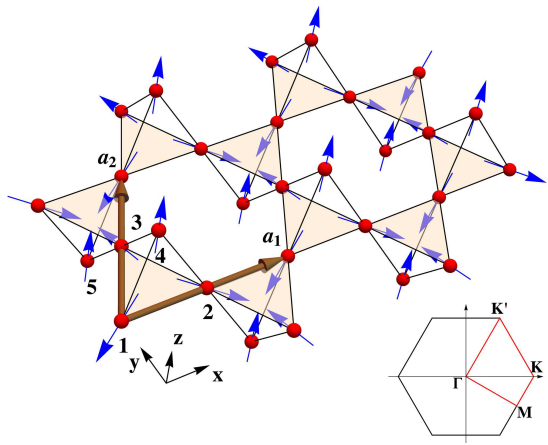}
\caption{\label{fig:Lattice} (Color online.) The triangular-kagome-triangular (TKT)  lattice structure with an all-in/all-out (AIAO) spin ordering for a [111] oriented growth. Red spheres and blue arrows are Ir atoms and the effective spins-1/2 degrees of freedom, respectively. The number of atoms in the unit cell is five (3 from the central kagome plane, and 1 each from the tip of an ``up" and ``down" pointing tetrahedron), in contrast to the bulk case which has 4 atoms in the unit cell: Labeled sites 4 (5) consist of a triangular plane above (below) the kagome plane. The lattice vectors are $\bm{a}_1=a(1, 0, 0)$ and $\bm{a}_2=a(\frac{1}{2}, \frac{\sqrt{3}}{2}, 0)$ with a lattice constant $a$. The Brillouin zone and the path along high symmetry points used in Fig.~\ref{fig:bands} are shown at the right corner.}
\end{figure}

\subsection{\label{sec2A}Spin Hamiltonian}
To study the magnon thermal transport, we focus on quasi-2D pryrochlore iridate thin films with non-magnetic A site ions grown in the [111] direction,\cite{hu2015first,hu2012topological,chen2015correlation,laurell2017topological} as shown in Fig.~\ref{fig:Lattice}.  The alternating triangular and kagome layers along the [111] axis preserves the spatial inversion symmetry of the lattice. Because of the absence of a mirror symmetry along the [111] direction, the number of atoms in the unit cell of this triangular-kagome-triangular (TKT) thin film is different from the bulk system. However, we assume that the spin Hamiltonian, which consists of only nearest-neighbor interactions, takes the same form as in the bulk case,\cite{wan2010electronic,elhajal2005ordering,lee2013magnetic}
\begin{equation}
    H=\sum_{\langle ij\rangle}[J\bm{S}_i\cdot\bm{S}_j+\bm{D}_{ij}\cdot\bm{S}_i\times\bm{S}_j+S_i^a\Gamma_{ij}^{ab}S_j^b],
    \label{Hamiltonian}
\end{equation}
where $\bm{S}_i(S_i^a)$ is the spin moment (component) on site $i$, and $J$ represents the antiferromagnetic Heisenberg coupling, $\bm{D}_{ij}$ is the Dzyaloshinskii-Moriya (DM) interaction on bond $ij$, and $\Gamma_{ij}^{ab}$ is the symmetric anisotropic exchange coupling tensor. With a large cubic crystal field from the oxygen octahedra surrounding each Ir$^{4+}$ ion along with a strong spin-orbit coupling (SOC), we take the magnetic moment of Ir$^{4+}$ as a effective spin 1/2 because of the splitting of the $t_{2g}$ orbitals into total angular momentum 1/2 (partially filled) and 3/2 (completely filled) manifolds in the strong SOC limit.\cite{kim2009phase,zhang2013effective}

Based on symmetry alone, the direction of the DM vectors in the TKT thin film cannot be completely determined from Moriya's rule.\cite{moriya1960anisotropic,moriya1960new} However, if we only include the DMI arising from nearest neighbors, the DM vector of each bond is parallel to the opposite bond allowed by the mirror symmetry of the Ir$^{4+}$ tetrahedron.\cite{keffer1962moriya}  In our study of the magnon Nernst effects from topological magnon bands, we choose the sign of the DM vectors to obtain a stable AIAO spin ordering,\cite{elhajal2005ordering} as shown in Fig.~\ref{fig:Lattice}. In such a state, the couplings can be parametrized as \cite{laurell2017topological} 
\begin{align}
J&=\frac{4t^2}{U}\left[\cos^2\left(\frac{\theta_t}{2}-\theta\right)-\frac{1}{3}\sin^2\left(\frac{\theta_t}{2}-\theta\right)\right],\\
\bm{D}_{ij}&=\frac{8t^2}{U}\cos\left(\frac{\theta_t}{2}-\theta\right)\sin\left(\frac{\theta_t}{2}-\theta\right)\hat{v}_{ij},\\
\Gamma_{ij}^{ab}&=\frac{8t^2}{U}\sin^2\left(\frac{\theta_t}{2}-\theta\right)\left(\hat{v}_{ij}^a\hat{v}_{ij}^b-\frac{\delta^{ab}}{3}\right),
\label{parametrization}
\end{align}
where $\theta_t=2\arctan\sqrt{2}$ is the tetrahedral angle, $t$ is the nearest-neighbor hopping energy, $U$ is the on-site Hubbard interaction, and $\theta$ is a parameter that controls the ratio among the couplings. The parameter $\theta$ depends on the material details,\cite{lee2013magnetic} and is not easy to freely control by static external parameters. However, out-of-equilibrium, with Floquet engineering, it is possible to control the interactions by using circularly polarized light. In such a periodically driven system, the Hubbard interaction $U$ can be renormalized into an effective term $U(\omega)$ that depends on the frequency $\omega$ and polarization of the light.\cite{bukov2016schrieffer,hejazi2019floquet} In addition, laser illumination can modify the hopping terms and Hubbard terms indirectly with lattice vibrations and distortions.\cite{forst2011nonlinear,nova2017effective} With these two methods, the coupling strengths can be experimentally tuned in principle, in addition to the application of static substrate strain and hydrostatic pressure which provide a more limited control route.

\subsection{\label{sec2B}Spin Wave Analysis}
To study the magnon Nernst effect, a spin wave analysis is necessary to obtain magnon dispersions.  Because of the non-collinearity of the system, there is no global $S_z$ direction, so one must orient the Cartesian coordinate system for each sublattice such that the $\bm{\hat{z}}$-axis locally lies along the classical ground-state orientation of the onsite macro-spins.\cite{owerre2017noncollinear,flebus2019interfacial,ma2020longitudinal} In other words, the spin $\bm{S}_i(\theta_i, \phi_i)$ is related to the one in the local frame of reference, $\bm{S}_{i}'$, as $\bm{S}_{i}' = \mathcal{R}_{i} \bm{S}_{i}$ with
\begin{eqnarray}
\mathcal{R}_{i}&=&\mathcal{R}_{y}(-\theta_{i}) \mathcal{R}_{z}(-\phi_{i})\nonumber\\
    &=&\left[\begin{array}{ccc}
\cos{\theta_i} & 0 & -\sin{\theta_i}\\
0                   & 1 & 0\\
\sin{\theta_i} & 0 & \cos{\theta_i}
\end{array}\right]\left[\begin{array}{ccc}
\cos{\phi_i}   & \sin{\phi_i} & 0\\
-\sin{\phi_i}  & \cos{\phi_i}   & 0\\
0                   & 0                   & 1
\end{array}\right],\label{Rotation}
\end{eqnarray}
where the matrix $\mathcal{R}_{z (y)}(\theta)$ is a right-handed rotational matrix of angle $\theta$ about the $\hat{\mathbf{z}}(\hat{\mathbf{y}})$ axis, and $\theta_{i} (\phi_{i})$ is the polar (azimuthal) angle of the classical ground-state orientation of $\mathbf{S}_{i}$.

In the local reference frame of each sublattice, a local $\bm{\hat{z}}$ direction is well-defined and the sublattice spin can then be expressed with a Holstein-Primakoff representation\cite{holstein1940field} 
\begin{equation}
\left\{\begin{array}{ll}
{S'}_{i}^+=\sqrt{2S-{a_i}^\dag a_{i}} a_{i},\\
{S'}_{i}^z=\left( S -{a_i}^\dag a_{i}\right),
\end{array}\right.\label{HP}
\end{equation}
where $S=\frac{1}{2}$ is the magnitude of the local spin.

If we ignore the higher order terms leading to magnon-magnon interactions, the spin Hamiltonian can be truncated to quadratic order as,
\begin{align}
    H=\frac{1}{2}\mathbf{X_k}^\dag H_{\mathbf{k}}\mathbf{X_k},\label{2nd_Quantization}
\end{align}
where $\mathbf{X_k}=(a_1(\mathbf{k}),\dots,a_5(\mathbf{k}),a_1^\dag(-\mathbf{k}),\dots,a_5^\dag(-\mathbf{k}))^T$ expands the Hilbert space into a particle-hole space (PHS) and $H_{\mathbf{k}}$ stands for a bosonic Bogoliubov-de Gennes (BdG) Hamiltonian,\cite{shindou2013topological} also in a particle-hole symmetric form as
\begin{align}
H_{\mathbf{k}}=\left[\begin{array}{cc}
A(\mathbf{k}) & B(\mathbf{k})\\
B^*(-\mathbf{k}) & A^*(-\mathbf{k})
\end{array}\right].\label{BdG}
\end{align}
To diagonalize this BdG Hamiltonian, one needs to use a paraunitary matrix $Q_{\mathbf{k}}$ which satisfies
\begin{align}
    Q^\dag_{\mathbf{k}}\sigma_3Q_{\mathbf{k}}=\sigma_3,\,\, Q^\dag_{\mathbf{k}}H_{\mathbf{k}}Q_{\mathbf{k}}=\left[\begin{array}{cc}
E_{\mathbf{k}} & \\
& E_{-\mathbf{k}}
\end{array}\right],\label{Bogoliubov}
\end{align}
where $\sigma_3=\text{Diag}(1,-1)\bigotimes\mathbb{I}_5$ denotes the bosonic commutator in particle-hole space, and $E_\mathbf{k}$ are the eigeneneries. Here, $Q_\mathbf{k}$ can be regarded as a general Bogoliubov transformation\cite{shindou2013topological,del2004quantum} similar to the case in collinear antiferromagnets.\cite{kittel1963quantum} (More details can be found in Appendix \ref{appa}.)

\begin{figure*}
\includegraphics[width=0.96\textwidth]{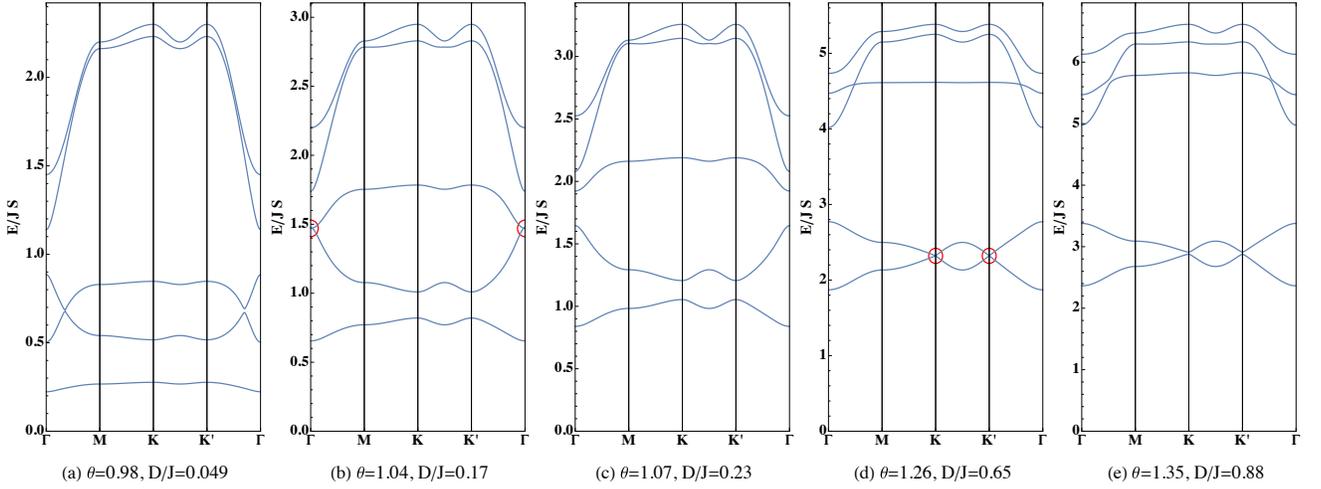}
\caption{\label{fig:bands}(Color online.) Spin-wave spectra for different $\theta$ values. The high symmetry path is shown in Fig.~\ref{fig:Lattice}. Energies are given in units of $JS$. The Chern numbers change when band gaps close and reopen at $\theta\approx1.04$ and $\theta\approx1.26$, as shown in (b) and (d) with red circles denoting the gap closings. The Chern numbers counting from the bottom are (a)$(+1, +1, -2, +1, -1)$, (c)$(+1, -1, -1, +2, -1)$ and (e)$(-1, +1, +3, -1, -2)$.}
\end{figure*}

With Eqs.~(\ref{Rotation})-(\ref{Bogoliubov}), we obtain the magnon band evolution by varying the ratio control parameter $\theta$.  We observe several gap closings and reopenings among the five bands (see Appendix \ref{appb} for more details), which changes the band topology (shown in Fig.~\ref{fig:bands}).   As the Chern numbers are different below and above the band gap, we expect magnon edge currents to carry spin angular momenta producing a magnon Nernst spin current in the presence of a temperature gradient.\cite{mook2014edge,ruckriegel2018bulk}

\subsection{\label{sec2C}Edge States}
\begin{figure}
\includegraphics[width=0.48\textwidth]{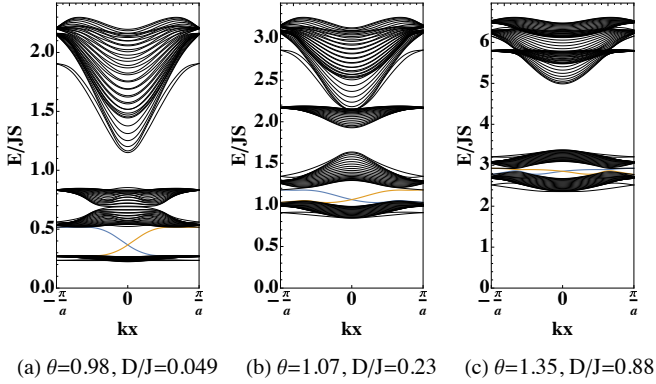}
\caption{\label{fig:ES} (Color online.) Magnon spectra of a 40 atom wide TKT thin film strip.   The strip is oriented along the $x$-direction with finite width in the $y$-direction. It is periodic along $x\ (\bm{a}_1)$-direction, following the convention in Fig.~\ref{fig:Lattice}. Blue and orange dispersions are the topologically protected magnon edge states spatially separated by the bulk (i.e., they are localized on different edges). Orange states are localized on the right, and blue states localized on the left of the strip.  Note the change in edge state direction in (c) relative to (a), (b).}
\end{figure}

\begin{figure}
\includegraphics[width=0.48\textwidth]{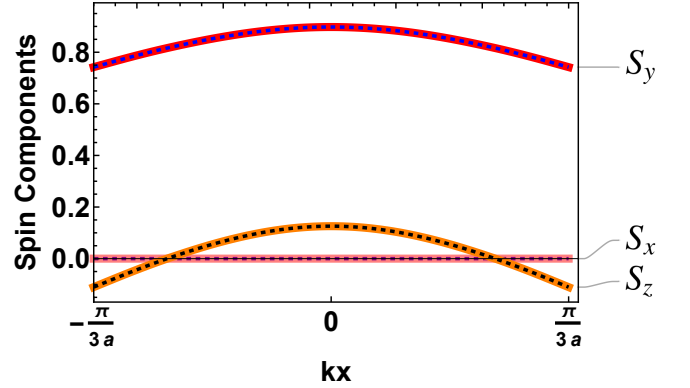}
\caption{\label{fig:S} (Color online.) Spin components of the edge modes using a strip of 40 atoms width and $\theta=1.07$. Solid (dashed) lines are for left (right) movers as shown in Fig.~\ref{fig:ES}. Only the $y$ and $z$-components are non-zero.}
\end{figure}

In contrast to previously studied systems with magnon Nernst effects,\cite{cheng2016spin,zyuzin2016magnon,li2020intrinsic} in our TKT thin films there is a direct gap above the lowest band. Since the sum of Chern numbers below a gap defines a winding number that is in direct correspondence with the number of edge modes,\cite{kim2016realization,mook2014edge} we further study the spin current at the edges in a strip geometry.  In Fig.~\ref{fig:ES}, there are two opposite edge modes (i.e., one on each edge but one being a continuation of the other in a finite area strip/system) within the gap as the lowest band has Chern number $+1(-1)$ for $\theta<(>)1.26$. The $\mathbf{k}$-dependence of the spin of these two edges is plotted in Fig.~\ref{fig:S} which shows that the chiral edge modes propagating along the two edges have the same $y$ and $z$ spin components due to the inversion symmetry between the pairs, and thus contribute opposite spin currents, while there is no $x$-component, which vanished due to the $M_{yz}\mathcal{T}$ symmetry of the system.

\section{\label{sec3}Magnon Nernst Effect}
In this section, we study a magnon Nernst effect, which can generate a transverse magnon spin current $\bm{J}^s$ in the TKT thin film from a longitudinal temperature gradient $\nabla T$. Here we only focus on the intrinsic effect of magnon Berry curvature, which induces an anomalous velocity and a transverse motion of magnons.\cite{matsumoto2011rotational} Generally, the spin angular momentum in a non-collinear system with SOC is not conserved.\cite{shi2006proper,nikolic2006imaging} In this case, one cannot denote a good spin quantum number for each magnon band and define a spin current. In our case, the non-collinearity arises from the DMI, which breaks the conservation of a spin angular momentum. However, with a spatial inversion symmetry, a total s-polarized spin current contribution from the MNE, $J^s_i=\alpha^s_{ij}\nabla_j T$, is well-defined\cite{li2020intrinsic} and the magnon Nernst coefficient (MNC) $\alpha^s_{ij}$ can be obtained as,
\begin{align}
    \alpha^s_{ij}=\frac{2k_B}{V}\sum_{n=1}^5\sum_{\mathbf{k}}(\Omega^s_{n,\mathbf{k}})_{ij}c_1[g(E_{n\mathbf{k}})],\label{MNC}
\end{align}
where $c_1(x)=(1+x)\ln(1+x)-x\ln x$, $g(x)=(e^{x/k_BT}-1)^{-1}$ is the Bose-Einstein distribution, and a spin Berry curvature is defined as, \cite{zyuzin2016magnon,li2020intrinsic}
\begin{equation}
    (\Omega^s_{n,\mathbf{k}})_{ij}=\sum_{m\neq n}(\sigma_3)_{nn}\frac{2\text{Im}[(j^s_{i\mathbf{k}})_{nm}(\sigma_3)_{mm}(v_{j\mathbf{k}})_{mn}]}{\left[(\sigma_3)_{nn}E_{n\mathbf{k}}-(\sigma_3)_{mm}E_{m\mathbf{k}}\right]^2},\label{sBC}
\end{equation}
where $v_{\beta\mathbf{k}}=\nabla_\beta H_\mathbf{k}$,  $j^s_{i\mathbf{k}}=\frac{1}{4}(v_{i\mathbf{k}}\sigma_3\hat{S}^s+\hat{S}^s\sigma_3v_{i\mathbf{k}})$ is the $s$-polarized spin current operator, and $(\dots)_{nm}$ stands for $Q^\dag_{n\mathbf{k}}(...)Q_{m\mathbf{k}}$ as matrix elements in the Bogoliubov representation.

Our TKT thin films are inversion symmetric so we can use Eq.~(\ref{sBC}) for calculating the spin current. More precisely, the thin film has a spatial point group $D_{3d}$ with generators $\{\mathcal{I}, M_{yz} ,C_{3z}\}$, and a magnetic point group $\bar{3}m'$ with the AIAO spin ordering. From a symmetry point of view, this magnetic point group is compatible with ferromagnetism, and thus a thermal Hall current is expected.\cite{mook2019thermal,suzuki2017cluster} Similarly, by considering the magnetic point group acting on the MNE tensor, one finds that there are only 4 individual response coefficients appearing in the tensors,\cite{aroyo2011crystallography}
\begin{align}
&\alpha^x=\left[\begin{array}{ccc}
-\alpha_1 & 0 & 0\\
0 & \alpha_1 & -\alpha_4\\
0 & -\alpha_3 & 0
\end{array}\right],\,\,
\alpha^y=\left[\begin{array}{ccc}
0 & \alpha_1 & \alpha_4\\
\alpha_1   & 0 & 0\\
\alpha_3 & 0 & 0
\end{array}\right],\nonumber\\
&\alpha^z=\left[\begin{array}{ccc}
0 & \alpha_2 & 0\\
-\alpha_2   & 0 & 0\\
0 & 0 & 0
\end{array}\right].\label{TKTMNC}
\end{align}

The structure of these tensors is consistent with our results for the edge states (Fig.~\ref{fig:S}), as $\alpha^x$ has no transverse coefficients while $y$ and $z$-polarized spin components can transport along edges transversely for a temperature gradient along the $y$-direction. From a more simple picture, since the total net moment within one unit cell is only along the out-of-plane direction, i.e. $z$-direction, $\alpha^x$ and $\alpha^y$ with in-plane polarizations should have similar independent coefficients, $\alpha_i$ for $i=1,3,4$, while $\alpha^z$ is different from them (containing only $\alpha_2$).

\begin{figure*}
\includegraphics[width=0.96\textwidth]{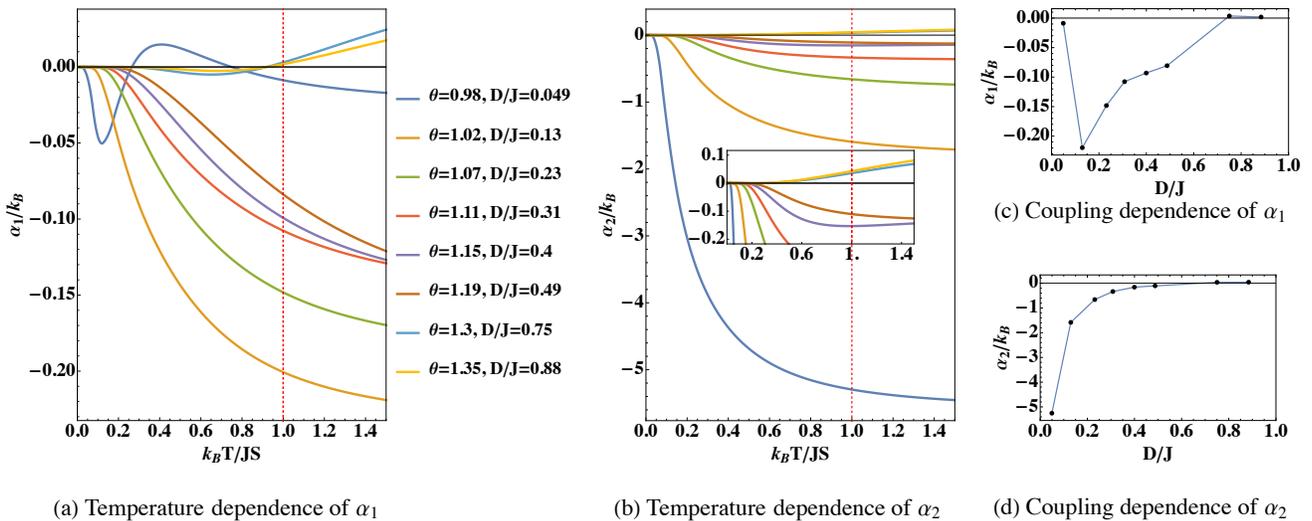}
\caption{\label{fig:MNC}(Color online.) (a-b) The temperature dependence  and (c-d) DMI dependence of the magnon Nernst coefficients. The inset in (b) is a zoom-in to show the sign change of $\alpha_2$ for $\theta>1.26$. (c) and (d) are plotted along the red dashed line in (a) and (b) respectively.}
\end{figure*}

The typical lattice constant between iridium ions of bulk pyrochlore iridates is on the order of 10\AA,\cite{shapiro2012structure} and thus in our thin film case, we focus on the response induced by a temperature gradient within the plane, e.g coefficient $\alpha_1$ and $\alpha_2$ corresponding to the magnon Nernst response $\alpha^y_{xy}$ and $\alpha^z_{xy}$. The temperature dependence of these coefficients is shown in Fig.~\ref{fig:MNC} 

\begin{figure*}
\includegraphics[width=0.477\textwidth]{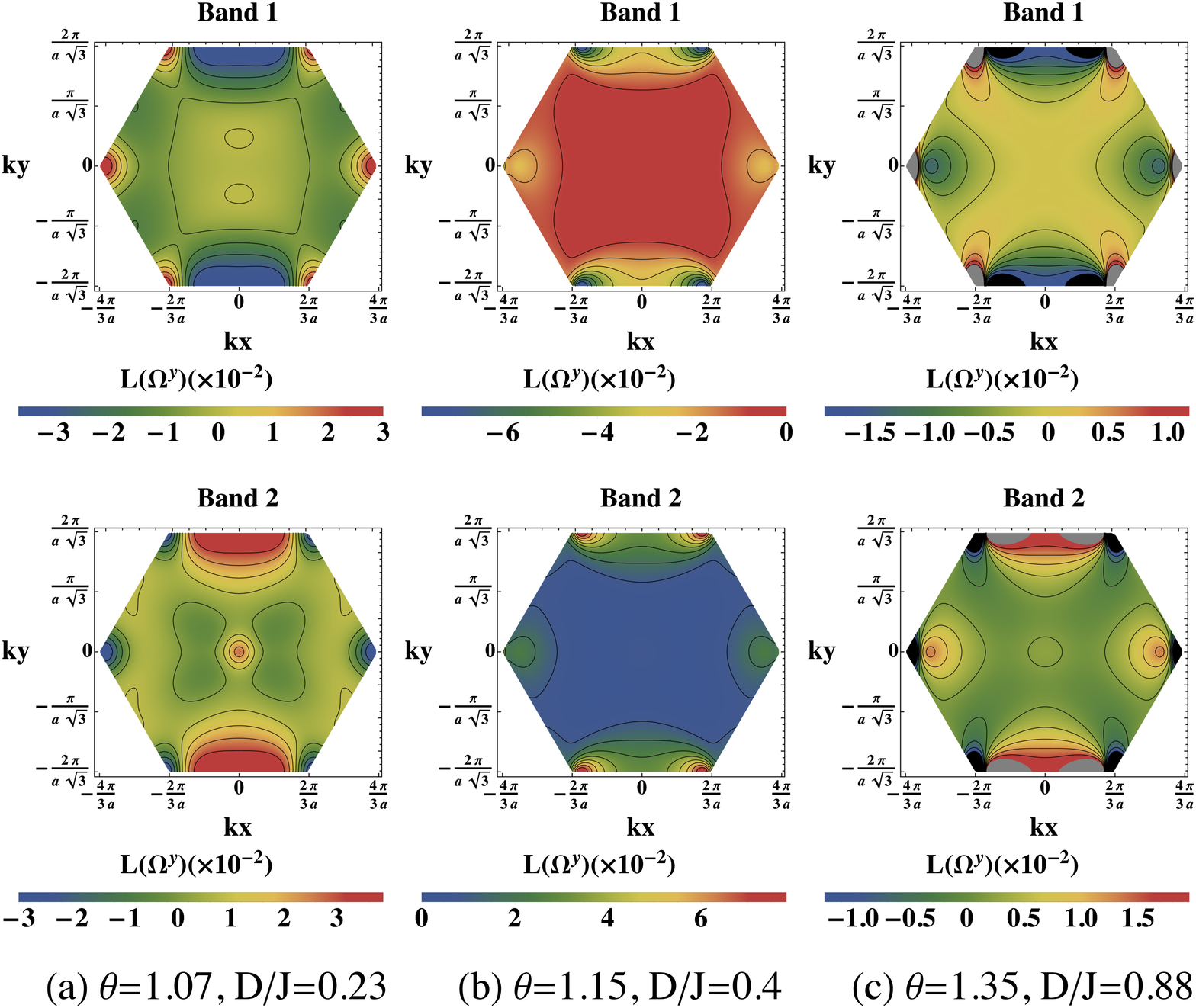}
\includegraphics[width=0.48\textwidth]{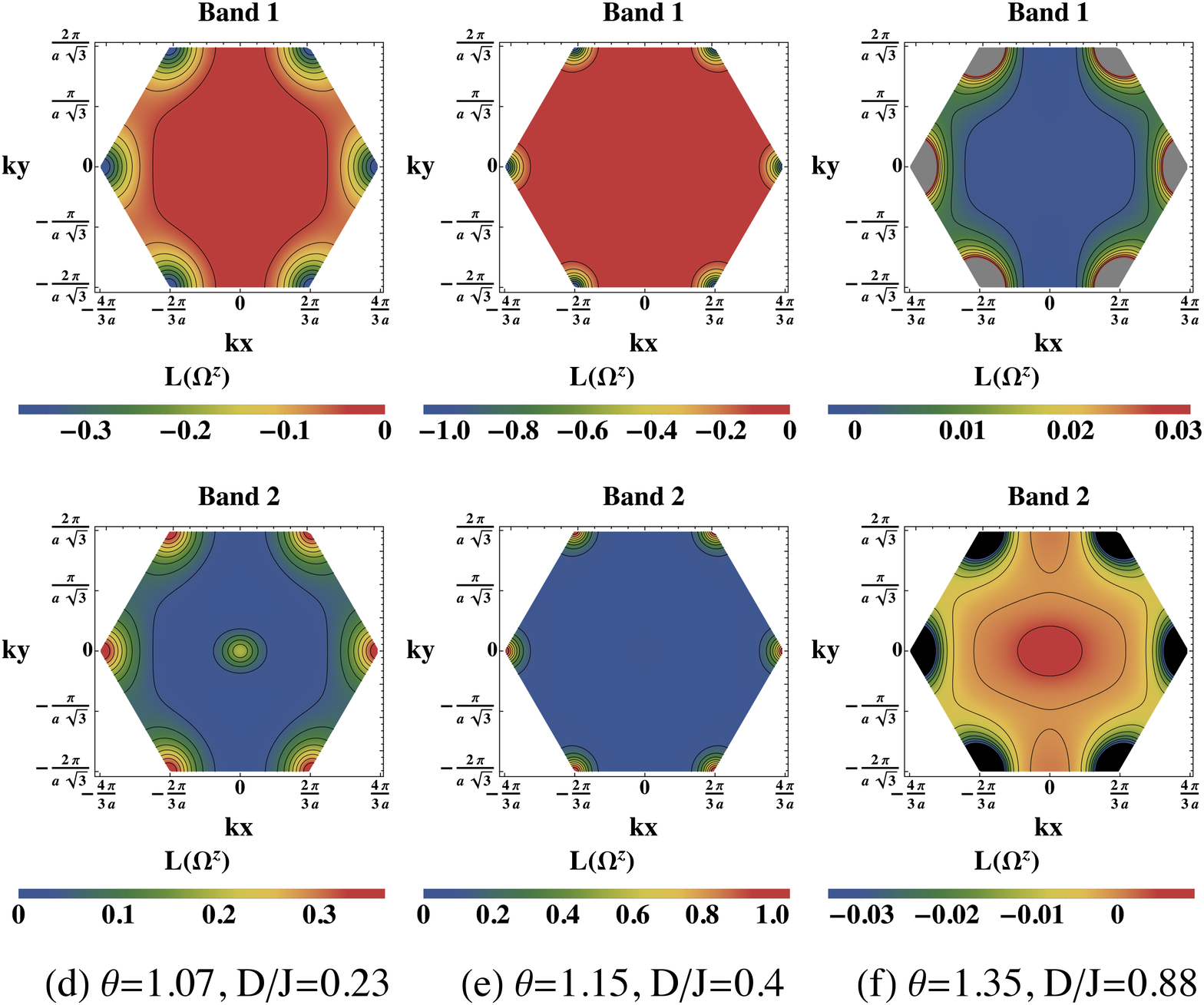}
\caption{\label{fig:sBC} (Color online.) Density plots of the spin Berry curvatures in log scale $L(\Omega^s)=\text{sign}(\Omega^s)\log(1+|\Omega^s|)$ for the lowest two bands (Band 1 is the lowest band). (a-c) are $y$-polarized spin Berry curvatures for different $\theta$ values before and after a gap closing. (d-f) are for $z$-polarized spin Berry curvatures. Both spin Berry curvatures concentrate mostly around the $\mathbf{K}$ and $\mathbf{K}'$ points with opposite signs. Detailed plots of the spin Berry curvature in the vicinity of the gray and black regions, corresponding to the values outside of the range of the scale bar, can be found in Appendix \ref{appc}, Fig.~\ref{fig:sBCapp}.}
\end{figure*}

From Fig.~\ref{fig:MNC}(c),(d), we can see that the DMI in general suppresses the response coefficients. On the one hand, from Fig.~\ref{fig:bands}, we have a higher excitation energy of the system with a larger DMI. Then the magnon bands can be accessed only with higher temperature as $c_1(g(E_{n\mathbf{k}}))$ decreases exponentially with increasing energy. On the other hand, most contributions to the MNE are from the lowest two bands, and as shown in Fig.~\ref{fig:sBC} the spin Berry curvature of these two bands concentrates at the $\mathbf{K}$ and $\mathbf{K}'$ points with opposite signs because of the DMI. When the DMI increases, the two bands move towards one another, and thus $\alpha^s\propto\Omega^s_1(\mathbf{K})(E_{1\mathbf{K}}-E_{2\mathbf{K}})$ is getting smaller. After the two bands touch each other, the MNC increases again as the gap reopens when $\theta>1.26$. 

In the high temperature limit, the MNC changes sign when $\theta\gtrsim 1.26$, due to the sign change of Chern numbers in the bottom two bands after a topological phase transition. However, since the net moment is along z-direction, $\alpha_2(\alpha^z_{xy})$ results from the spin angular momentum carried by a total thermal Hall magnon current, while $\alpha_1(\alpha^y_{xy})$ comes from the imbalance among magnon modes similar to the magnon Nernst effect in collinear antiferromagnets.\cite{cheng2016spin} Because of this, $\alpha_1$ is more sensitive to the band topology and change of the DMI. 

For $\theta=0.98$, the DMI is small compared to the exchange coupling and the system approaches the Heisenberg limit. The low energy scale of this limit supports the contributions to the MNE from higher bands when the temperature increases. Thus, instead of a monotonic change with respect to the temperature, $\alpha_1$ changes sign at $k_BT\approx0.27JS$ and $k_BT\approx0.74JS$ reflecting the Chern numbers with alternative sign from the bottom to top as (+1, +1, -2, +1, -1). 

For $\theta>1.26$, although the Chern number of the lowest band changes sign to +1, the spin Berry curvature around the $M$-point with lower energy dominates at low temperature [see Fig.~\ref{fig:sBC}(c)] and thus gives rise to a response with a negative sign. When temperature increases, the sign of response coefficient will change to positive as the change of Chern number during the phase transition, which is determined by the (spin) Berry curvature at two concentrated points $K$ and $K'$.

\section{\label{sec4}Discussion and Conclusion}
We have calculated the magnon Nernst response coefficients in the TKT pyrochlore iridates thin film with all-in/all-out spin ordering. In contrast with the magnon Nernst effect in collinear systems, we found transverse spin currents with both $y$- and $z$-polarization (for a thermal gradient applied along the $x$-direction). As an experimental estimate, if we assume a thickness of 20 \AA\ (for a trilayer, taken from the lattice constant of a material\cite{shapiro2012structure,shinaoka2015phase} such as Y$_2$Ir$_2$O$_7$) with a 20 K/mm temperature gradient, we expect a $y$-polarized ($z$-polarized) spin current on the order of $10^{-11} (10^{-10})$ J/m$^2$, which makes pyrochlore iridates a promising experimental candidate for observing the magnon Nernst effect in non-colliner systems.

In addition, we found the TKT trilayer film has a direct gap which supports two topological nontrivial edge modes. These edge modes have not been discussed in the context of the magnon Nernst effect before. They provide a spin current channel at each edge with opposite sign, and may be detected separately at the edge by local measurements.\cite{du2017control} Since these edge modes are topologically protected, it is possible to realize a spin current only weakly affected by disorder in the bulk.\cite{xu2006stability}

Our results also show that the DMI can significantly impact the response coefficients, as it can modify nontrivial geometric aspect of magnon bands, including topological transitions. The sign change of the coefficients is due to the distribution of spin Berry curvature throughout the Brillouin zone and the thermal population of  bands with different Chern numbers. In order to observe the sign change of $\alpha_1$ for $\theta=0.98\ (1.30)$, one must have an experimental resolution of the order of $10^{-11}\ (10^{-12})$ J/m$^2$, which can be converted into and measured by an inverse spin Hall current on the order of $10^{4}\ (10^{3})$ A/m$^2$.\cite{sinova2015spin} Therefore, the magnon Nernst effect can also be used as a probe to study the topological properties of magnons in magnetic insulators.  In addition, the magnon spin transport can be tuned with an external magnetic field, either by a spin ground state phase transition in the strong field limit or a Zeeman energy splitting in the weak field limit.\cite{ueda2017magnetic,li2016weyl}

\begin{acknowledgments}
We thank Pontus Laurell and Benedetta Flebus for discussions on spin wave analysis. BM also thanks Nemin Wei and Naichao Hu for helpful discussions on magnon band topolgy, and Martin Rodriguez-Vega for suggestion on tuning DM interactions. We gratefully acknowledge support from NSF DMR-1949701 and NSF DMR-2114825, with additional support from the NSF through the Center for Dynamics and Control of Materials: an NSF MRSEC under Cooperative Agreement No. DMR-1720595. This work was performed in part at the Aspen Center for Physics, which is supported by National Science Foundation grant PHY-1607611.
\end{acknowledgments}

\appendix

\section{\label{appa}GENERAL BOGOLIUBOV TRANSFORMATION}
As described in Sec.~\ref{sec2A}, in order to obtain a physical energy spectra, one needs to use a paraunitary matrix $Q$ to diagonalize a BdG Hamiltonian. More generally, for a bosonic BdG Hamiltonian 
\begin{align}
    H=\mathbf{X}^\dag \bm{H}\mathbf{X}\label{HX},
\end{align}
with a basis $\mathbf{X}$ satisfying a commutator relation
\begin{align}
    [\mathbf{X}, \mathbf{X}^\dag]=g,\label{gX}
\end{align}
it can be transformed into a bosonic representation $\mathbf{Y}=Q^{-1}\mathbf{X}$ which has diagonalized spectra as 
\begin{align}
    H=\mathbf{Y}^\dag \bm{E}\mathbf{Y},\label{HY}
\end{align}
and a standard bosonic commutator in particle-hole space as
\begin{align}
    [\mathbf{Y}, \mathbf{Y}^\dag]=\sigma_3,\label{gY}
\end{align}
where $\sigma_3=\text{Diag}(1,-1)\bigotimes\mathbb{I}_{N\times N}$ and $\bm{E}$ is a diagonal matrix.
By comparing Eq.~(\ref{HX})(\ref{gX}) with Eq.~(\ref{HY})(\ref{gY}), it can be seen that
\begin{align}
    g=Q[\mathbf{Y}, \mathbf{Y}^\dag]Q^\dag=Q\sigma_3Q^\dag\Rightarrow Q^\dag=\sigma_3Q^{-1}g\label{gXY},
\end{align}
and
\begin{align}
    Q^\dag\bm{H}Q=\bm{E}\Rightarrow Q^\dag=Q^{-1}g\bm{H}Q=\sigma_3\bm{E}.
\end{align}
If there is no dispersionless Goldstone mode in the system, $\det{g\bm{H}}\neq0$ and the eigenvectors $P$ of $g\bm{H}$ can be found easily in any numerical methods, and if there is no degeneracy (or the degeneracy can be avoided in numerics), the eigenvectors can be rearranged so that
\begin{align}
    Q=PT,\label{Q=PT}
\end{align}
where $T=\text{Diag}(t_1,t_2,\dots,t_{2N})$.
Taking Eq.~(\ref{Q=PT}) into Eq.~(\ref{gXY}), one finds
\begin{align}
    P^{-1}g(P^\dag)^{-1}=T\sigma_3T^\dag=\sigma_3 \text{Diag}(|t_1|^2,|t_2|^2,\dots,|t_{2N}|^2)\label{T},
\end{align}
is a diagonal matrix and $|t_i|$ can be solved from Eq.~(\ref{T}).
Therefore, the paraunitary matrix $Q$ can be constructed from eigenvectors $P$ as
\begin{align}
    Q=P(P^\dag g^{-1}P\sigma_3)^{-\frac{1}{2}}U,
\end{align}
where $U$ is a $U(1)$ phase factor that can be chosen as the identity.

In our magnonic system, $g=\sigma_3$, but in principle this general Bogoliubov transformation can be applied to any bosonic system such as phonons\cite{zhang2015chiral} and hybrid bosons.\cite{zhang2019thermal,zhang20203,go2019topological}
\section{\label{appb}GAP CLOSING AND CHERN NUMBERS}
\begin{figure}
\includegraphics[width=0.48\textwidth]{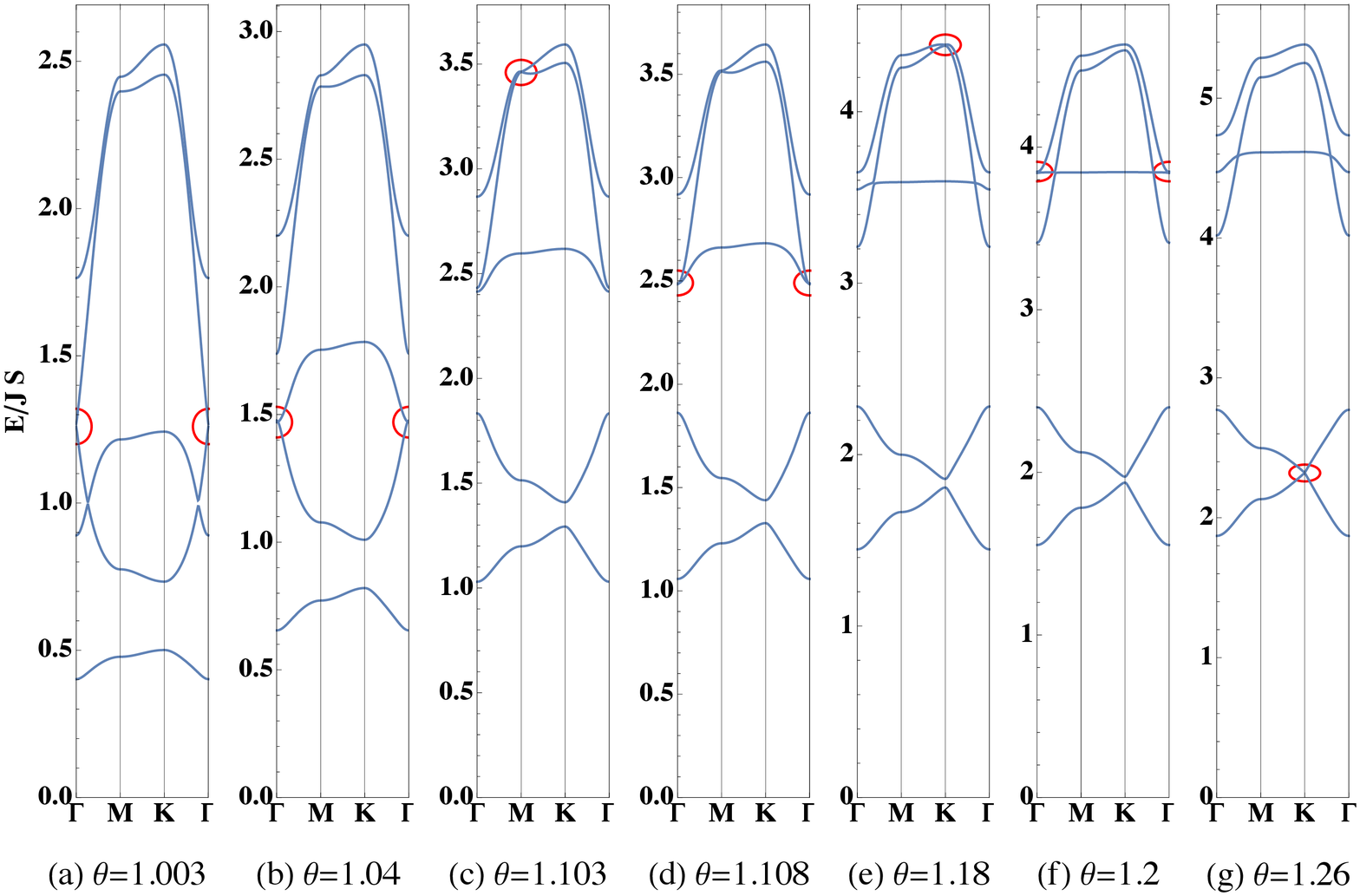}
\caption{\label{fig:gaps} (Color online.) Magnon band touching points, at which there occur topological phase transitions. Red circles identify the touching points.}
\end{figure}

When $\theta$ increases from $\arctan\sqrt{2}$ to $\pi-\arctan\sqrt{2}$, the DMI becomes more dominant in the Hamiltonian and the magnon bands touch and reopen several times, which separates the system into different topological phases. In Fig.~\ref{fig:gaps}, we show all band touching points and we list the Chern numbers in the phases between touchings in Table~\ref{tab:Chern_Num}.

\begin{table*}
\caption{\label{tab:Chern_Num}Chern numbers in the phases between touchings. We calculate the Chern numbers with the Fukui lattice discretization method\cite{takahiro2013chern} for momentum space grids of $501\times501$ sites. }
\begin{ruledtabular}
\begin{tabular}{ccccccccc}
$\theta$ & $<1.003$ & 1.003 - 1.04 & 1.04 - 1.103 & 1.103 - 1.108 & 1.108 - 1.18 & 1.18 - 1.20 & 1.20 - 1.26 & $>1.26$ \\
$D/J$ & $<0.096$ & 0.096 - 0.17 & 0.17 - 0.3 & 0.3-0.31 & 0.31-0.47 & 0.47 - 0.51 & 0.51 - 0.65 & $>0.65$\\
\hline
Bottom band & +1 & +1 & +1 & +1 & +1 & +1 & +1 & -1\\
 &+2&+2&-1&-1&-1&-1&-1&+1\\
 &-3&-4&-1&-1&-3&+3&+3&+3\\
 &+1&+2&+2&-1&+1&-3&-1&-1\\
Top band &-1&-1&-1&+2&+2&0&-2&-2\\
\end{tabular}
\end{ruledtabular}
\end{table*}

\section{\label{appc}SPIN BERRY CURVATURE AROUND HIGH VALUE POINTS}
\begin{figure}
\includegraphics[width=0.48\textwidth]{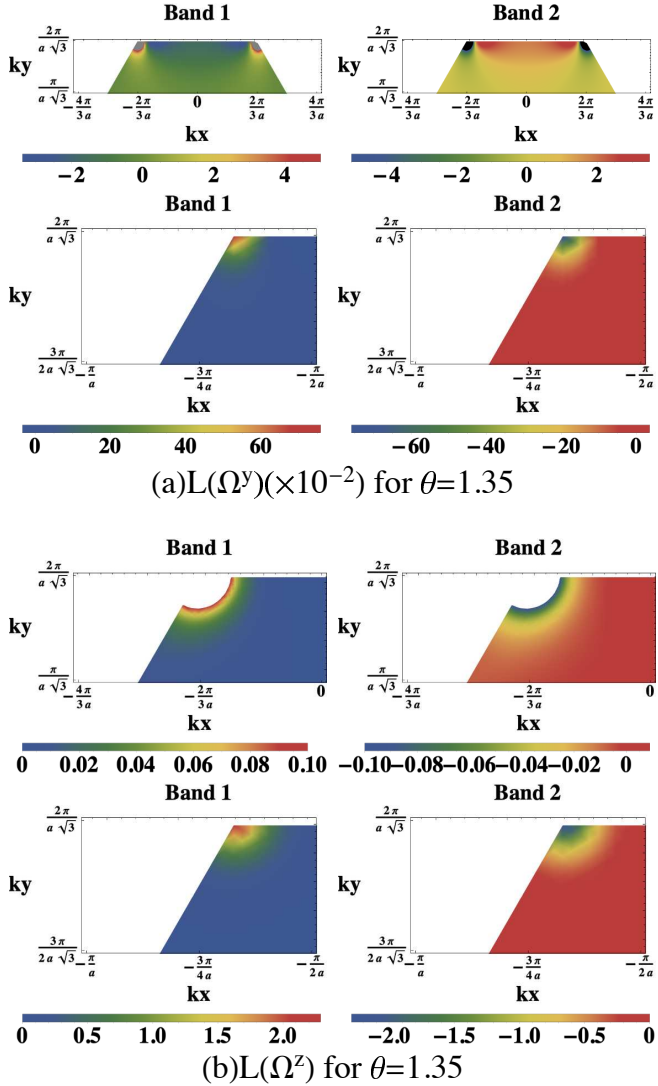}
\caption{\label{fig:sBCapp} (Color online.) Spin Berry curvature in different ranges around $\mathbf{K}$, $\mathbf{K'}$ and $\mathbf{M}$ points.}
\end{figure}

One can see in Fig.~\ref{fig:sBCapp} that in band 1 (2), there are negative (positive) spin Berry curvatures around $\mathbf{M}$ with lower energies, while a larger positive (negative) spin Berry curvature concentrates at $\mathbf{K}$ and $\mathbf{K'}$ points. The sign change of $\alpha_1$ can be explained as a competion between them.

\nocite{*}

\bibliography{PI_SNE}

\end{document}